\documentclass[a4paper,12pt]{article}
\usepackage[top=1.25in, bottom=1.25in, left=1.05in, right=1.05in]{geometry}\usepackage[utf8]{inputenc}
\usepackage[T1]{fontenc}
\usepackage[english]{babel}

\usepackage{natbib}

\usepackage[section]{placeins}

\usepackage{setspace}

\usepackage{enumitem}
\setlist{noitemsep}  % Reduce space between list items (itemize, enumerate, etc.)

\usepackage[colorinlistoftodos,color=orange!60]{todonotes}
\usepackage{regexpatch}
\makeatletter
\xpatchcmd{\@todo}{\setkeys{todonotes}{#1}}{\setkeys{todonotes}{inline,#1}}{}{}
\makeatother

\usepackage{amscd,amsmath,amssymb,amsfonts,amsthm,bbm,bm,latexsym,mathrsfs,mathtools,bigints}
\usepackage[subnum]{cases}	% multi-case equations with separate number for each
\usepackage{dsfont}
\makeatletter
\newcommand*{\distas}[1]{\mathbin{\overset{#1}{\kern\z@\sim}}}	
\makeatother
\newcommand*\abs[1]{\left|#1\right|}		% absolute value
\allowdisplaybreaks

\newtheorem{proposition}{Proposition}%[section]
\theoremstyle{remark}

\theoremstyle{plain}

\RequirePackage[colorlinks=true, citecolor=blue, urlcolor=blue]{hyperref}

\usepackage{xcolor,subfigure,epsfig,epstopdf,rotating,graphics,graphicx} 
\graphicspath{{./figures/}}
\usepackage[width=0.9\textwidth, font={footnotesize}]{caption}

\usepackage{rotating,lscape,pdflscape}
\usepackage{booktabs,longtable,float,array,multirow,colortbl,adjustbox}

\newcolumntype{C}[1]{>{\centering\arraybackslash}p{#1}}

\makeatother

\makeatletter
\def \Y{\mathbf{Y}}
\def \y{\mathbf{y}}

\def \btheta{\boldsymbol{\theta}}
\def \bepsilon{\boldsymbol{\epsilon}}

\def \b{\mathbf{b}}
\def \B{\mathbf{B}}

\def \bSigma{\boldsymbol{\Sigma}}

\def \R {\mathds{R}}

\def \E {\mathbb{E}}
\def \MAL{\textnormal{MAL}}
\def\S {\mathds{S}_{++}}

\makeatother

\title{\vspace{-60pt} \textbf{Bayesian Mixed-Frequency Quantile Vector Autoregression: Eliciting tail risks of Monthly US GDP}
}

\author{
Matteo Iacopini\thanks{Queen Mary University of London, United Kingdom. \color{blue}\texttt{m.iacopini@qmul.ac.uk}}
\and
Aubrey Poon\thanks{\"Orebro University, Sweden. \color{blue}\texttt{aubrey.poon@oru.se}}
\and
Luca Rossini\thanks{University of Milan, Italy. \color{blue}\texttt{luca.rossini@unimi.it}}
\and
Dan Zhu\thanks{Monash University, Australia. \color{blue}\texttt{dan.zhu@monash.edu}}
}

\date{\today}

\begin{document}

\maketitle

\begin{abstract}
Timely characterizations of risks in economic and financial systems play an essential role in both economic policy and private sector decisions.
However, the informational content of low-frequency variables and the results from conditional mean models provide only limited evidence to investigate this problem.
We propose a novel mixed-frequency quantile vector autoregression (MF-QVAR) model to address this issue. Inspired by the univariate Bayesian quantile regression literature, the multivariate asymmetric Laplace distribution is exploited under the Bayesian framework to form the likelihood. A data augmentation approach coupled with a precision sampler efficiently estimates the missing low-frequency variables at higher frequencies under the state-space representation.

The proposed methods allow us to nowcast conditional quantiles for multiple variables of interest and to derive quantile-related risk measures at high frequency, thus enabling timely policy interventions.
The main application of the model is to nowcast conditional quantiles of the US GDP, which is strictly related to the quantification of Value-at-Risk and the Expected Shortfall.

\end{abstract}

\clearpage

%\onehalfspacing
\doublespacing

\section{Introduction}

Most economic models' primary object of interest is the conditional mean of a given variable or index, as it summarizes the central response to explanatory variables.
However, following the financial crises and economic shocks that characterized the last decade, policymakers and researchers have shifted their attention and interest beyond the conditional mean.
In particular, the significant effects that exogenous shocks (such as the COVID-19 pandemic), wars (e.g., the Russian-Ukrainian war), and fluctuations of the business cycles have on the economy highlight the crucial need to investigate the tail and shoulders of the response variable's distribution.

Timely characterizations of risks to the economic outlook play a vital role in both economic policy and private sector decisions, where central bankers and analysts share a demand for timely forecasts of economic activity. To encapsulate and reflect the most recent events, forecasts of macroeconomic or financial variables should blend information collected from a wide array of sources and observed at different intervals or frequencies.

Moreover, research following the global financial crisis has provided substantial empirical evidence that the relationships among macroeconomic and financial time series are characterized by nonlinearities and asymmetries \citep{hubrich2015financial,kilian2017role,adrian2019vulnerable}.
Thus, investigating the nonlinear effects related to cycles is crucial to policymakers for designing policies targeted at specific phases of the cycles.
Researchers in macroeconomics usually base their analysis on linear regression methods, whereas only recently have nonlinear methods been applied to investigate economic policies and financial crises \citep[e.g.,][]{caggiano2017uncertainty,Huber2022BART}.
However, using conditional mean regression methods raises several concerns when modeling data with features such as skewness, fat tails, outliers, truncation, censoring and heteroscedasticity. This relates to the fact that the impact of covariates on the response may significantly vary across the range of the latter, thus highlighting the limitations of methods based on conditional mean only. The issue is exacerbated by nonlinear relationships and non-Gaussian noises, typical features of many economic and financial variables.

In detecting economic and financial crises, appropriate risk measures are needed \citep{Merlo2021VaR}, such as the Value at Risk (VaR) or the Expected Shortfall (ES).
The VaR considers the maximum loss an operator can incur over a defined time horizon and for a given confidence level. At the same time, the ES coincides with the conditional expectation of exceedance beyond the VaR. % Whilst the VaR does not account for tail risk.
In forecasting, \cite{Gneiting2011qCRPS} propose a threshold- and quantile-based decomposition of the continuously ranked probability score to assess density forecasting over the whole distribution and specific quantiles or regions of a variable of interest (e.g., the tails).

%%%%%%%-------- OUR CONTRIBUTION --------%%%%%%%

We introduce a novel mixed-frequency quantile vector autoregressive (MF-QVAR) model to address these issues. Unlike standard linear regression models, quantile regression \citep[QR, see][]{koenker1978regression} provides robust modeling of conditional quantiles. It allows the covariates to exert different impacts on each quantile level, thus enabling a comprehensive investigation of the entire conditional distribution. Following \cite{petrella2019joint}, we rely on the multivariate asymmetric Laplace (MAL) distribution to form the likelihood and conduct a simultaneous inference under the Bayesian framework on the marginal conditional quantiles of a multivariate response variable, taking into account the possible correlation among the marginals. Our framework permits the investigation of asymmetry in the downside and upside risks by considering different quantile levels, unlike standard models with symmetric second-moment dynamics. This quantile regression approach can be more effective, especially when skewness dynamics accompany the evolution of the distribution.

We build on the mixed frequency literature to exploit the information available at a higher frequency to nowcast quantiles of low-frequency variables of interest based on the state-space representation. The nowcasting and forecasting of multivariate quantiles would enable the policymakers to promptly detect early signals of distress and adopt corrective measures to counteract the early deterioration of the system.
The proposed approach automatically allows the forecaster to overcome the differences in data release dates that cause the available information set to differ over time within the quarter, so-called the ``ragged-edge'' problem.

%\begin{figure}[t!h]
%\centering
%\hspace*{-3ex}
%\setlength{\abovecaptionskip}{1pt}
%\begin{tabular}{cc}
%\includegraphics[trim= 0mm 0mm 0mm 0mm,clip,height= 3.5cm, width= 7.5cm]{./data/data_GDP_inflation.eps} &
%\includegraphics[trim= 0mm 0mm 0mm 0mm,clip,height= 3.5cm, width= 7.5cm]{./data/data_GDP_inflation_zoom.eps}
%\end{tabular}
%\caption{Monthly inflation (grey, left axis) and quarterly GDP growth (black, right axis).}
%\end{figure}

To overcome the computational challenge in the mixed frequency VAR, we follow  \cite{chan2021efficient}, who designed a computationally efficient sampler for state-space models with missing observations, such as MF-VARs. They exploit the block-banded structure of the precision matrix of the conditional distribution of the missing observations to adapt the precision-based sampler of \cite{chan2009efficient} to draw the missing low-frequency variables \citep[see also][]{rue2001fast,rue2005gaussian}.
The importance of the precision sampler to modern econometric models is proven by its use in a variety of settings, including macro-econometric \citep{chan2013new}, models with missing observations \citep{hauber2021precision}, and dynamic factor models \citep{kaufmann2019bayesian}.
An earlier method to make inference on unobserved variables in state space models is the simulation smoother \citep{durbin2002simple,durbin2012time}.
Moreover, we impose linear constraints when sampling the missing observations to ensure that missing high-frequency observations match the observed values of the low-frequency variables. This results in sampling from a linearly constrained Gaussian distribution, which is efficiently performed following the methods in \cite{cong2017fast}.

We apply our novel MF-QVAR model on a real-time nowcasting application for monthly US growth-at-risk. Specifically, we focus on the out-of-sample evaluation period between January 2016 and March 2022. During this evaluation period, we encounter ragged edges at the end of the sample as we respect the release calendar for all the monthly and quarterly variables included in our MF-QVAR model. In particular, we focus on generating the monthly nowcasts and forecasts under the three release timings of the US real GDP. 

We found two key insights from our real-time nowcasting application. First, we show that there has been a downward shift in the monthly nowcasts of US growth-at-risk since the pandemic. For instance, the monthly nowcasts of US growth-at-risk were, on average, about -3\%, and this average dropped to about -5\% during the pandemic period. Thus, this result implies that the monthly distribution of US real GDP has become more skewed to the left since the pandemic. Second, we compare our monthly nowcasts of US growth-at-risk to their corresponding quarterly nowcasts from the \cite{adrian2019vulnerable} quantile regression model. We found that, on average, the quarterly nowcasts underestimate US growth-at-risk relative to our monthly nowcasts. Furthermore, our monthly nowcasts of US growth-at-risk appear to align with the current post-pandemic economic situation in the US. 

We also undertake a counterfactual analysis to investigate whether the Chicago Fed's National Financial Condition Index (NFCI) is important for nowcasting monthly US growth-at-risk. We compare our counterfactual monthly nowcasts to their corresponding actual monthly nowcasts. We found, on average, that a tightening of the NFCI does indeed have a negative impact on the monthly nowcast US growth-at-risk. Therefore, our results reinforce the findings of \cite{adrian2019vulnerable} and highlight the importance of including NFCI in a mixed frequency setting when modeling growth-at-risk.

The remainder of this article is organized as follows: Section~\ref{sec:model} presents a novel mixed-frequency quantile VAR model under the acronym of MF-QVAR and the intertemporal constraints.
In Section~\ref{sec:inference}, the Bayesian approach for inference along with the posterior algorithm is described.
% Section~\ref{sec:simulations} investigates the performance of our method using simulated data.
Section~\ref{sec:application} shows the results of a real data and a counterfactual analysis on US real GDP growth-at-risk.
Finally, Section~\ref{sec:conclusion} draws the conclusions.

\section{Mixed Frequency Quantile VAR}    \label{sec:model}

\subsection{Notation}
Let $\mathds{S}^k = \{ X \in \R^{k\times k} : X=X' \}$ denote the space of symmetric matrices of size $k\times k$ and $\S^k = \{ X \in \mathds{S}^k : \mathbf{a}'X\mathbf{a} >0, \; \forall \, \mathbf{a} \in \R^k \}$ be the space of symmetric, positive definite matrices of size $k\times k$.
For a matrix $A\in\S^k$, $A^{1/2}$ represents the Cholesky factor of $A$.
Let $\MAL_n(\boldsymbol{\mu},\boldsymbol{\delta},\Sigma)$ indicate a multivariate asymmetric Laplace distribution with location $\boldsymbol{\mu} \in\R^n$, skewness parameter $\boldsymbol{\delta} \in\R^n$, and scale matrix $\Sigma \in\S^n$.
Let $\y_t^o \in\R^{n_o}$ be a vector of variables observed at high frequency and let $\y_t^u \in\R^{n_u}$ be a vector of variables that are unobserved or only partially observed.
Finally, let us denote with $I_n$ the identity matrix of size $n$, and use $\mathbf{1}_n$ and $\mathbf{0}_n$ to represent an $n$-dimensional vector with all entries equal to one and zero, respectively. The symbol $\otimes$ denotes the Kronecker product.

\subsection{Model}
We consider a $n$-dimensional VAR($p$) model with $p$ lags for $\y_t = (\y_t^{o^\prime}, \y_y^{u^\prime})'$, with $n=n_o+n_u$, that is
\begin{equation}
\y_{t} = \b_0 + \sum_{j=1}^{p} B_{j} \y_{t-j} + \bepsilon_{t}, \qquad
\bepsilon_{t} \sim \MAL_n(\mathbf{0}_n, D\btheta_{\tau,1}, D\btheta_{\tau,2} \Psi \btheta_{\tau,2}' D'),
\label{eq:VARp_MAL}
\end{equation}
for $t=p+1,\ldots,T$, where $\b_0$ is a $n$-dimensional vector of intercepts, $B_1,\ldots,B_p$ are $(n\times n)$ autoregressive coefficient matrices, $\Psi$ is a $(n\times n)$ correlation matrix, $D = \operatorname{diag}\big( \Sigma_{11}^{1/2},\ldots,\Sigma_{nn}^{1/2} \big)$, for $\Sigma_{ii}^{1/2} \in\R$, and 
\begin{align*}
\btheta_{\tau,1} & = \left( \frac{1-2\tau_{1}}{\tau_{1}(1-\tau_{1})}, \ldots, \frac{1-2\tau_{n}}{\tau_{n}(1-\tau_{n})} \right)', \qquad
\btheta_{\tau,2} = \operatorname{diag} \left( \sqrt{\frac{2}{\tau_{1}(1-\tau_{1})}}, \ldots, \sqrt{\frac{2}{\tau_{n}(1-\tau_{n})}} \ \right),
\end{align*}
where $\bm{\tau} = (\tau_1,\ldots,\tau_n)$ is the quantile representation, such that $\tau_i \in (0,1)$ for $i=1,\ldots,n$.
 
The multivariate asymmetric Laplace distribution, $\MAL_n(\boldsymbol{\mu}, D\btheta_{\tau,1}, D\btheta_{\tau,2} \Psi \btheta_{\tau,2}' D')$, has density function
\begin{align*}
f_Y(\mathbf{y} | \boldsymbol{\mu}, D\btheta_{\tau,1}, D\btheta_{\tau,2} \Psi \btheta_{\tau,2}' D') &= \frac{2\exp{\left\{(\mathbf{y} -\boldsymbol{\mu})'D^{-1} (\btheta_{\tau,2} \Psi \btheta_{\tau,2}')^{-1} \btheta_{\tau,1}\right\}}}{(2\pi)^{n/2} |D\btheta_{\tau,2} \Psi \btheta_{\tau,2}' D|^{1/2}} \left(\frac{\tilde{m}}{2+\tilde{d}}\right)^{\nu} K_{\nu} \left(\sqrt{(2+\tilde{d}) \tilde{m}}\right),
\end{align*}
where $\tilde{m} = (\mathbf{y} - \boldsymbol{\mu})' (D\btheta_{\tau,2} \Psi \btheta_{\tau,2}' D')^{-1} (\mathbf{y} - \boldsymbol{\mu})$, $\tilde{d} = \btheta_{\tau,1}'\btheta_{\tau,2} \Psi \btheta_{\tau,2}' \btheta_{\tau,1}$ and $K_{\nu}(\cdot)$ denotes the modified Bessel function of the third kind with index parameter $\nu = (2-n)/2$. The MAL distribution  is closely related to multivariate quantile regression models, as stated in Proposition~\ref{prop:MAL_petrella} from \cite{petrella2019joint}, which we report using our notation.
\begin{proposition}[\cite{petrella2019joint}] \label{prop:MAL_petrella}
	Let $\mathbf{y} \sim \MAL_n(\boldsymbol{\mu}, D\btheta_{\tau,1}, D\btheta_{\tau,2} \Psi \btheta_{\tau,2}' D')$ and let $\boldsymbol{\tau} = (\tau_1,\ldots,\tau_n)'$ be a fixed $n$-dimensional vector, such that $\tau_i \in (0,1)$ for $i=1,\ldots,n$. Then $\mathbb{P}(y_i \leq \mu_i) = \tau_i$ if and only if
	\begin{equation*}
	\theta_{\tau,1,i}   = \frac{1-2\tau_i}{\tau_i(1-\tau_i)}, \qquad 
	\theta_{\tau,2,i}^2 = \frac{2}{\tau_i(1-\tau_i)}.
	\end{equation*}
	Moreover, $y_i \sim \mathcal{AL}(\mu_i,\Sigma_{ii}^{1/2},\tau_i)$ follows a univariate asymmetric Laplace distribution (see Appendix~\ref{sec:apdx_asymmetric_Laplace}).
\end{proposition}

Modern models for investigating macroeconomic and financial time series are inherently multivariate, which calls for developing suitable multivariate quantile regression models.
\cite{Manganelli2021quantileIRF} provide a structural quantile VAR (QVAR) model to capture nonlinear relationships among macroeconomic variables, and propose a quantile impulse response function to perform stress tests. \cite{MontesRojas2019MultQIRF} develop a reduced form QVAR model to provide reliable forecasts and define a different quantile impulse response function to explore dynamic heterogeneity of the response variables to exogenous shocks.
Recently, \cite{adams2021forecasting} modified the approach of \cite{adrian2019vulnerable} and used quantile regressions to characterize upside and downside risks around the survey of professional forecasters' median consensus forecasts for each indicator.

The above-mentioned studies adopt a frequentist perspective. In contrast, \cite{bernardi2015bayesian} developed a Bayesian inference for univariate quantile regression models to measure tail risk interdependence using \cite{tobias2016covar}'s Conditional VaR (CoVaR) indicator, defined as a quantile of a conditional distribution calculated at a given quantile of its conditioning distribution.  Recently, \cite{tian2021bayesian} exploited shrinkage priors to estimate Bayesian multivariate quantile regressions. 
Despite the increasing interest by policymakers in understanding and forecasting the whole distribution of economic and financial indicators, the literature on quantile VAR models is scant.
We aim to fill this gap by proposing a novel fast Bayesian approach to inference for quantile VAR models.

%  Xt = (n x n^2) * p --> n x n^2p
Following the literature on multivariate time series, we define the $n_\beta$-dimensional vector $\boldsymbol{\beta} = (\b_0', \operatorname{vec}(B_1)',\ldots, \operatorname{vec}(B_p)')'$, with $n_\beta = n(1+np)$, and the $n \times n_\beta$-dimensional matrix $X_t = (\mathbf{1}_n,\mathbf{x}_{t,1}',\ldots,\mathbf{x}_{t,p}')$, with $\mathbf{x}_{t,j} = (\mathbf{y}_{t-j} \otimes I_n)$ for each $j=1,\ldots,p$.
Moreover, let us reparametrize the innovation scale by introducing the positive definite matrix $\Sigma = D \Psi D' \in \S^n$, and relabelling with $D = D(\Sigma) = \operatorname{diag}\big( \Sigma_{11}^{1/2},\ldots,\Sigma_{nn}^{1/2})$.
Owing to the properties of the multivariate asymmetric Laplace distribution, eq.~\eqref{eq:VARp_MAL} admits a representation as a location-scale mixture of Gaussian distributions \citep{petrella2019joint,kotz2001laplace}, as follows:
\begin{align}
\label{eq:VARp}
\y_{t} & = \b_0 + \sum_{j=1}^{p} B_{j} \y_{t-j} + D\btheta_{\tau,1} w_t + \sqrt{w_t} D\btheta_{\tau,2} \Psi^{1/2} \tilde{\mathbf{z}}_{t}, \qquad  \tilde{\mathbf{z}}_{t} \sim \mathcal{N}_n(\mathbf{0}_n, I_n), \\
       & = X_t \boldsymbol{\beta} + D(\Sigma)\btheta_{\tau,1} w_t + \mathbf{z}_{t}, \qquad  \mathbf{z}_{t} \sim \mathcal{N}_n(\mathbf{0}_n, w_t \btheta_{\tau,2} \Sigma \btheta_{\tau,2}),
\end{align}
where $w_t$ is an auxiliary variable satisfying\footnote{We use the rate parametrization, such that if $x \sim \mathcal{E}xp(1)$, then $kx \sim \mathcal{E}xp(1/k)$, for $k>0$.} $w_t \distas{i.i.d.} \mathcal{E}xp(1)$, and define  $\mathbf{w} = (w_{p+1},\ldots,w_{T})'$ the vector of latent variables.

\subsection{Mixed Frequency and Inter-Temporal Constraints}
Mixed frequency VAR (MF-VAR) models in macroeconomics and forecasting have become increasingly popular for producing high-frequency nowcasts of low-frequency variables. Specifically, MF-VARs are often used to joint model quarterly macroeconomic variables (low-frequency), such as gross domestic product (GDP), and monthly financial variables (high-frequency), such as surveys, to produce monthly nowcasts of GDP \citep[e.g., see][]{schorfheide2015real}.

We contribute to this literature by introducing mixed-frequency components in quantile VAR models that allow us to nowcast the conditional mean and the distribution of the low-frequency variables of interest. This is of paramount importance to timely understand the status of the economic system by explaining and nowcasting fundamental indicators, such as GDP and systemic risk indices. Moreover, it allows the researchers to consider any ragged-edge issues arising from the data release calendar.

A stacked or a state-space approach can be used to handle the mixed-frequency variables.
The first class includes the MIDAS models, initially proposed by \cite{ghysels2005there,ghysels2006predicting} and \cite{ghysels2016macroeconomics}, which consist of a linear model for the lowest observed frequency that includes the high-frequency covariates using particular functional forms for the coefficients.
This framework has recently been extended to account for multiple regimes, dynamic panels, and high-dimensional settings \citep[e.g., see][]{casarin2018uncertainty,khalaf2021dynamic,mogliani2021bayesian}. Conversely, the state-space approach proposed by \cite{schorfheide2015real} treats the high-frequency observations of the low-frequency variables as missing values and estimates them via Kalman filtering and smoothing algorithms. This method has been applied to investigate consumption growth and long-run risks \citep{schorfheide2018identifying}, the COVID-19 outbreak \citep{huber2020nowcasting}, the output gap \citep{cimadomo2021nowcasting}, regional output growth \citep{koop2020regional}, and high-dimensional macroeconomic systems \citep{berger2020nowcasting}.

In particular, we adopt the state-space approach and model all variables at the highest observed frequency to obtain the interpolated estimates of the low-frequency variables at a higher frequency.
The main drawback of this approach is the significant computational burden mainly due to the estimation of high-dimensional latent state vectors (i.e., missing observations of the low-frequency variables) via filtering and smoothing techniques. This cost becomes prohibitive as the dimension of the VAR gets large, thus representing a major obstacle to using state-space methods on datasets with medium-high dimensions.

To investigate the joint distribution of the unobserved variables, conditional on the observations, let us denote with $\Y = (\y_{p+1}',\ldots,\y_T')'$ and $\boldsymbol{\zeta} = (\mathbf{z}_{p+1}',\ldots,\mathbf{z}_T')'$ the $(T-p)n$-dimensional vectors obtained by stacking all observations and all innovations over time, respectively.
It is now possible to rewrite eq.~\eqref{eq:VARp} in matrix form as:
\begin{equation}
\B \Y = \b + \boldsymbol{\zeta}, \qquad \boldsymbol{\zeta} \sim \mathcal{N}(\mathbf{0}_{(T-p)n}, \bSigma),
\label{eq:VARp_matrix_form}
\end{equation}
where
\begin{align*}
\b & = (\mathbf{1}_{T-p} \otimes \b_0) + \mathbf{w} \otimes \left( \frac{(1-2\tau_{1})\Sigma_{11}^{1/2}}{\tau_{1}(1-\tau_{1})}, \ldots, \frac{(1-2\tau_{n})\Sigma_{nn}^{1/2}}{\tau_{n}(1-\tau_{n})} \right)', \quad \quad
\bSigma = \operatorname{diag}(\mathbf{w}) \otimes \btheta_{\tau,2} \Sigma \btheta_{\tau,2},
\end{align*}
and 
\begin{align*}
\B = \left[ \begin{array}{cccccccccc}
-B_{1} & -B_{2} & \ldots & -B_{p} & I_{n} & \mathbf{0}_{n} & \ldots & \mathbf{0}_{n} & \mathbf{0}_{n} & \mathbf{0}_{n} \\
\mathbf{0}_{n} & -B_{1} & \ldots & -B_{p-1} & B_{p} & I_{n} & \ldots & \mathbf{0}_{n} & \mathbf{0}_{n} & \mathbf{0}_{n} \\
\ldots & \ldots & \ldots & \ldots & \ldots & \ldots & \ldots & \ldots & \ldots & \ldots \\
\mathbf{0}_{n} & \mathbf{0}_{n} & \ldots & \mathbf{0}_{n} & \mathbf{0}_{n} & \mathbf{0}_{n} & \ldots & -B_{p} & I_{n} & \mathbf{\mathbf{0}}_{n}\\
\mathbf{0}_{n} & \mathbf{0}_{n} & \ldots & \mathbf{0}_{n} & \mathbf{0}_{n} & \mathbf{0}_{n} & \mathbf{0}_{n} & -B_{p-1} & -B_{p} & I_{n}
\end{array} \right]
\end{align*}
as a banded matrix of dimensionality $(Tn \times (T-p)n)$.
Notice that $\b$, $\bSigma$, and $\B$ depend on the model parameters; we drop the notation for exposition purposes. 

As stated in \cite{schorfheide2015real}, we formulate the mixed frequency VAR in a state-space structure.
To this aim, let us first denote with $\y^o = (\y_1^{o^\prime},\ldots,\y_T^{o^\prime})'$ and $\y^u = (\y_1^{u^\prime},\ldots,\y_T^{u^\prime})'$ the $Tn_o$- and $Tn_u$-dimensional stacked vectors of observed and unobserved response variables. This allows to represent the vector $\y$ as a linear combination of $\y^o$ and $\y^u$, as follows:
\begin{equation}
\y = M_{u} \y^{u} + M_{o} \y^{o},
\label{eq:linear_combination_Y}
\end{equation}
where $M_o$ and $M_u$ are $(Tn \times Tn_o)$ and $(Tn \times Tn_u)$ selection matrices with full column rank.
Then, by substituting eq.~\eqref{eq:linear_combination_Y} into eq.~\eqref{eq:VARp_matrix_form} along the lines of \cite{chan2021efficient}, one obtains the joint distribution of the missing observations, conditional on the observed data and model parameters as the Gaussian distribution:
\begin{align}
\y^{u} | \y^{o},\bm{\beta},\Sigma,\mathbf{w}  & \sim \mathcal{N}(\boldsymbol{\mu}_{y}, K_{y}^{-1}),
\label{eq:distribution_yu_given_yo}
\end{align}
where $K_{y} = M_{u}' \B' \bSigma^{-1} \B M_{u}$ and $\boldsymbol{\mu}_{y} = K_{y}^{-1} \left( M_{u}' \B' \bSigma^{-1} \left( \b - \B M_{o} \y^{o} \right) \right)$.

In practice, the unobserved data points in $\y^u$ are constrained to match the value of the low-frequency variables at those points in time when the latter are observed.
One of the most commonly used restrictions for log-differenced variables is the log-linear approximation of \cite{mariano2003new,mariano2010coincident}.
This approach assumes that the observed quarterly value of the $i$-th variable at month $t$, denoted by $\tilde{y}^u_{i,t}$, is obtained as a linear combination of the missing monthly values at the current and previous four months, denoted by $y_{i,t}^u,\ldots,y_{i,t-4}^u$, as follows:
\begin{equation}
\tilde{y}_{i,t}^u = \frac{1}{3} y_{i,t}^u + \frac{2}{3} y_{i,t-1}^u + y_{i,t-2}^u + \frac{2}{3}y_{i,t-3}^u + \frac{1}{3} y_{i,t-4}^u.
\label{eq:intertemporal_constraint}
\end{equation}
This is a log-linear approximation to an arithmetic average of the quarterly variable, where note that $\tilde{y}^u_{i,t}$ is only observed for every third month.
Stacking the inter-temporal constraints over time, one gets
\begin{equation}
\tilde{\y}^{u} = M_{a} \y^{u},
\label{eq:intertemporal_constraint_matrix_form}
\end{equation}
where $M_{a}$ is a $(k \times Tn_u)$ matrix containing the $k$ linear restrictions, and $\tilde{\y}^{u}$ is a vector containing the observed values of the low-frequency variables.
To account for the inter-temporal constraints when sampling the unobserved variables, $\y^u$, it is sufficient to draw from the Gaussian distribution in eq.~\eqref{eq:distribution_yu_given_yo}  subject to the restrictions in eq.~\eqref{eq:intertemporal_constraint_matrix_form}.
This is efficiently done following the methods described in Algorithm 2 of \cite{cong2017fast}, which postulates first to draw a vector from the unconstrained distribution, $\mathbf{u} \sim \mathcal{N}(\boldsymbol{\mu}_y, K_y^{-1})$, then compute
\begin{equation}
\y^{u} = \mathbf{u} + K_{y}^{-1} M_{a}' (M_{a} K_{y}^{-1} M_{a}')^{-1} (\tilde{\y}^{u} -M_{a} \mathbf{u}).
\label{eq:sample_yu_givn_yo_intertemporal}
\end{equation}
From a computational perspective, we follow the efficient implementation in Algorithm 1 of \cite{chan2021efficient}.

\section{Bayesian inference}    \label{sec:inference}

In this section, we provide the details of the estimation of our proposed MF-QVAR model. Initially, we exploit the location-scale mixture representation of the multivariate asymmetric Laplace in eq.~\eqref{eq:VARp_MAL} and introduce a set of auxiliary variables $w_t \distas{i.i.d.} \mathcal{E}xp(1)$, thus the complete-data likelihood is given by:
\begin{align}
L(&(\textbf{y}_1^o,\textbf{y}_1^u),\ldots, (\textbf{y}_T^o,\textbf{y}_T^u),\mathbf{w} | \bm{\beta},\Sigma)  = \prod_{t=p+1}^T P(\y_t | \bm{\beta},\Sigma,w_t) P(w_t | \bm{\beta},\Sigma) \notag \\
 & = \prod_{t=p+1}^T \exp\Big\{ -\frac{1}{2}\Big[ \big( \y_t - X_t \boldsymbol{\beta} - D(\Sigma) \btheta_{\tau,1} \big)' \big( w_t \btheta_{\tau,2} \Sigma \btheta_{\tau,2} \big)^{-1} \big( \y_t - X_t \boldsymbol{\beta} - D(\Sigma) \btheta_{\tau,1} \big) \Big] \Big\} \notag \\
 & \quad \times (2\pi)^{-\frac{n}{2}} \abs{w_t \btheta_{\tau,2} \Sigma \btheta_{\tau,2}}^{-\frac{n}{2}} \exp\{ -w_t \}.
\label{eq:complete_likelihood}
\end{align}
Before describing the posterior distributions along with the algorithm used, we define the prior specifications for our parameters. Starting with the coefficient vector $\boldsymbol{\beta} = (\b_0', \operatorname{vec}(B_1,\ldots,B_p)')'$,  we assume a  
 conjugate multivariate Gaussian prior distribution
\begin{align*}
\boldsymbol{\beta} \sim \mathcal{N}_{n_\beta}(\boldsymbol{\underline{\mu}}_\beta, \underline{\Omega}_\beta).
\end{align*}
For this vector of coefficients, one may consider using shrinkage priors such as the global-local shrinkage prior \citep{polson2010shrink,bhadra2016default}, and the Minnesota prior \citep{kadiyala1997numerical}. In this article, we focus on a simple case by setting the prior mean of the coefficient associated with each equation at the frequentist univariate regression estimate, $\underline{\boldsymbol{\mu}}_\beta = \widehat{\boldsymbol{\beta}}$, and a prior variance $\underline{\Omega}_\beta = 100 \cdot I_{n_\beta}$, which results in a relatively flat prior distribution. We leave for further research the use of more complex shrinkage priors.

The other parameter of interest is the scale matrix, $\Sigma \in \S^n$, and in this scenario, we assume an inverse Wishart prior distribution
\begin{equation*}
\Sigma \sim \mathcal{IW}_n(\underline{\nu}_0, \ \underline{\Phi}_0),
\end{equation*}
where $\underline{\nu}_0 > n-1$ is the degrees of freedom parameter and $\underline{\Phi}_0 \in\S^n$ is a scale matrix, such that if $\underline{\nu}_0 > n+1$ then $\E[\Sigma] = \underline{\Phi}_0 / (\underline{\nu}_0-n-1)$.
This is equivalent to assuming the Wishart prior distribution for $\Sigma^{-1} \sim \mathcal{W}_n(\underline{\nu}_0, \ \underline{\Phi}_0^{-1})$, where $\E[\Sigma^{-1}] = \underline{\Phi}_0^{-1} \underline{\nu}_0$.

%To deal with the overfitting issues that affect medium and large VAR models, one may consider a global-local shrinkage prior for $\boldsymbol{\beta}$ \citep{polson2010shrink,bhadra2016default}.
%Commonly-used choices include the horseshoe \citep{carvalho2010horseshoe}, the Dirichlet-Laplace \citep{bhattacharya2015dirichlet}, and the Normal-Gamma \citep{brown2010inference} priors, which have been recently applied to Bayesian VAR models \citep{follett2019achieving,kastner2020sparse,huber2019adaptive}.
%This class of prior distributions is obtained by assuming $\underline{\boldsymbol{\mu}}_\beta = \mathbf{0}_{n_\beta}$, $\underline{\Omega}_\beta = \operatorname{diag}(\omega_{1},\ldots,\omega_{n_\beta})$, and specifying a hierarchical prior for the variance of each element $\beta_j$ as
%\begin{align*}
%\beta_j | \lambda_j \psi & \sim \mathcal{N}(0, \lambda_j \psi), \qquad
%\lambda_j \sim \pi(\lambda_j), \qquad
%\psi \sim \pi(\psi),
%\end{align*}
%where $\lambda_j$ and $\psi$ are the local and global components \citep{polson2010shrink}.

%\subsection{Posterior Sampling}

Based on these prior specifications and the likelihood function in eq.~\eqref{eq:complete_likelihood}, we can provide the full conditional distributions for each parameter and latent variable of the model. 
We remark that our parametrization differs from \cite{tian2016bayesian}, as we work with the positive definite matrix $\Sigma$, instead of the correlation matrix $\Psi$ and the diagonal matrix $D$ separately.
Moreover, we work with multivariate Bayesian analysis of quantile regression models, while \cite{yu2001bayesian} proposed a Bayesian approach for the univariate framework.

As the joint posterior distribution is not tractable, we rely on data augmentation to obtain closed-form full conditional distributions and to design an efficient Markov chain Monte Carlo (MCMC) algorithm for approximating the posterior distribution.
Specifically, we design an efficient Gibbs sampler based on the precision sampler of \cite{chan2009efficient} and \cite{chan2021efficient}, which cycles over the following steps:
\begin{enumerate}
\item draw $\y^u$ given $\textbf{y}^o, \bm{\beta}, \textbf{w}$ and $\Sigma$ from eq.~\eqref{eq:distribution_yu_given_yo} subject to the restrictions in eq.~\eqref{eq:intertemporal_constraint_matrix_form} using Algorithm 1 of \cite{chan2021efficient};
\item draw $\boldsymbol{\beta}$ given $\textbf{y}^o, \textbf{y}^u, \textbf{w}$ and $\Sigma$ from the Gaussian distribution $\mathcal{N}_{n_\beta}(\boldsymbol{\overline{\mu}}_b, \overline{\Omega}_b)$, with $\mathbf{\tilde{e}}_t = \y_t - D\btheta_{\tau,1} w_t$ and parameters 
\begin{align*}
\overline{\Omega}_{b} & = \Big( \underline{\Omega}_b^{-1} + \!\!\!\sum_{t=p+1}^T \!\! X_t' (w_t \btheta_{\tau,2} \Sigma \btheta_{\tau,2})^{-1} X_t \Big)^{-1}, \quad
\boldsymbol{\overline{\mu}}_{b}  = \overline{\Omega}_{b}^{-1} \Big( \underline{\Omega}_b^{-1} \boldsymbol{\underline{\mu}}_b + \!\!\!\sum_{t=p+1}^T \! \mathbf{\tilde{e}}_t' (w_t \btheta_{\tau,2} \Sigma \btheta_{\tau,2})^{-1} X_t \Big).
%
%\overline{\Omega}_{b} & = (\btheta_{\tau,2}^{-1} \Sigma^{-1} \btheta_{\tau,2}^{-1}) \otimes (X' \operatorname{diag}(\mathbf{w})^{-1} X) + \underline{\Omega}_b^{-1} \\
%\boldsymbol{\overline{\mu}}_{b} & = \overline{\Omega}_{b}^{-1} \left(\operatorname{vec}\left(X' \operatorname{diag}(\mathbf{w})^{-1} (\Y - \btheta_{\tau,1} D \mathbf{w}) \btheta_{\tau,2}^{-1} \Sigma^{-1} \btheta_{\tau,2}^{-1} \right) + \underline{\Omega}_b^{-1} \boldsymbol{\underline{\mu}}_b \right);
\end{align*}

\item draw the auxiliary variables $w_t$, for each $t=p+1,\ldots,T$, given $\textbf{y}^o, \textbf{y}^u, \bm{\beta}$ and $\Sigma$ from the Generalized inverse Gaussian distribution $\text{GiG}(\overline{p}_w, \overline{a}_w, \overline{b}_{w,t})$, with $\mathbf{u}_t = \y_t -X_t\boldsymbol{\beta}$, and
\begin{align*}
\overline{p}_{w} = 1-\frac{n}{2}, \qquad
\overline{a}_{w} = 2 + \btheta_{\tau,1}'D (\btheta_{\tau,2} \Sigma \btheta_{\tau,2})^{-1} D \btheta_{\tau,1}, \qquad
\overline{b}_{w,t}  = \mathbf{u}_t' (\btheta_{\tau,2} \Sigma \btheta_{\tau,2})^{-1} \mathbf{u}_t.
%\overline{a}_{w} & = (D \btheta_{\tau,1})' \btheta_{\tau,2}^{-1} \Sigma^{-1} \btheta_{\tau,2}^{-1} (D \btheta_{\tau,1}) \\
%\overline{b}_{w,t} & = (\y_t -I_n \otimes X_t' \boldsymbol{\beta})' \btheta_{\tau,2}^{-1} \Sigma^{-1} \btheta_{\tau,2}^{-1} (\y_t -I_n \otimes X_t' \boldsymbol{\beta});
\end{align*}

\item draw $\Sigma$ given $\textbf{y}^o, \textbf{y}^u, \bm{\beta}$ and $\textbf{w}$ via the slice sampling algorithm of \cite{Neal2003slicesampling}. Defining $\mathbf{e}_t(\Sigma) = \y_t - X_t \boldsymbol{\beta} -D(\Sigma) \btheta_{\tau,1} w_t$, the target density function is proportional to:
\begin{align*}
%p(\Sigma | \bullet) 
 \propto \abs{\Sigma}^{-\frac{\underline{\nu}_0+n+1}{2}} \exp\left\{ -\frac{1}{2} \operatorname{tr}\big( \underline{\Phi}_0 \Sigma^{-1} \big) \right\} \abs{\Sigma}^{-\frac{T}{2}} \exp\left\{ -\frac{1}{2} \sum_{t=1}^{T} \mathbf{e}_t'(\Sigma) \big( w_t^{-1} \btheta_{\tau,2}^{-1} \Sigma^{-1} \btheta_{\tau,2}^{-1} \big) \mathbf{e}_t(\Sigma) \right\}.
\end{align*}
\end{enumerate}
The Supplementary Material provides a detailed description of the MCMC algorithm along with the derivation of the full conditional distributions.

\section{Nowcasting Monthly US GDP growth-at-risk}  \label{sec:application}
We illustrate the utility of our proposed MF-QVAR model by undertaking a real-time nowcasting application for the growth-at-risk of US real GDP. To the best of our knowledge, this is the first study in the literature that explicitly nowcasts a monthly growth-at-risk estimate for US real GDP, whereas all the previous studies only modeled it at a quarterly frequency.
A key advantage of our MF-QVAR model is that it \textit{naturally} allows the forecaster to consider any ragged-edge issues arising from the data release calendar. 

We estimate an MF-QVAR model consisting of the quarterly US real GDP and eight monthly variables. Seven of the monthly variables included in our model are broadly similar to the monthly variables chosen in \cite{schorfheide2015real} standard MF-VAR model, whereas the last one is the NFCI.
This choice is motivated by the recent study of \cite{adrian2019vulnerable}, which showed that a tightening of the NFCI could lead to a large increase in the growth-at-risk for US real GDP.
Table~\ref{tab:Data-information} reports the details of each data variable and their respective transformations. All data vintages were gathered from the St. Louis ALFRED database. 

We undertake our real-time nowcasting application between January 2016 and March 2022, and we focus on generating the nowcasts and forecasts under three release timings of the US real GDP. 
Furthermore, we follow an expanding window approach where our initial training sample (or hold out) period is from January 1973 to December 2015. As our application is a real-time exercise, we respect the release calendar for monthly and quarterly variables; therefore, we face ragged edges at the end of the sample, mainly due to the release delay of real GDP relative to the monthly indicators.

\begin{table}[t!h]
\footnotesize
\centering
\begin{tabular}{lccc}
\toprule 
Description & Fred Mnemonic & Frequency & Transformation \tabularnewline
\midrule
\midrule 
Average Weekly Hours & AWHMAN & Monthly & $0.1\times x_{t}$ \tabularnewline
\midrule 
CPI Inflation & CPIAUCSL & Monthly & $100\bigtriangleup\text{ln}x_{t}$ \tabularnewline
\midrule 
Industrial Production & INDPRO & Monthly & $100\bigtriangleup\text{ln}x_{t}$ \tabularnewline
\midrule 
S\&P 500 & S\&P500 & Monthly & $100\bigtriangleup\text{ln}x_{t}$ \tabularnewline
\midrule 
Federal Funds Rate & FEDFUNDS & Monthly & Level \tabularnewline
\midrule 
10 years Government Treasury yield & GS10 & Monthly & Level \tabularnewline
\midrule 
Unemployment Rate & UNRATE & Monthly & Level \tabularnewline
\midrule 
Chicago Fed National Financial Condition Index & NFCI & Monthly & Level \tabularnewline
\midrule 
Real Gross Domestic Product  & GDPC1 & Quarterly & $400\bigtriangleup\text{ln}x_{t}$\tabularnewline
\bottomrule
\end{tabular}
\caption{Data information.}
\label{tab:Data-information}
\end{table}

\begin{table}[t!h]
\footnotesize
\resizebox{1.0\textwidth}{!}{
\centering
\begin{tabular}{ccc}
\toprule 
Description & Ragged-edge at the end of sample & Nowcast Classification \tabularnewline
\midrule
\midrule 
US Real GDP has no release delay & No & Forecast \tabularnewline
\midrule 
US Real GDP has a release delay of one month & Yes & Nowcast $T+1$ \tabularnewline
\midrule 
US Real GDP has a release delay of two months & Yes & Nowcast $T+2$ \tabularnewline
\bottomrule
\end{tabular}{\small\par}
}
\caption{Classifications of Nowcasts.}
\label{tab:Classifications-of-Nowcasts}
\end{table}

Table~\ref{tab:Classifications-of-Nowcasts} describes the three specific types of nowcasts that we produce from our model according to the release delay of the US real GDP. In particular, exploiting the proposed quantile regression framework, we focus on the growth-at-risk nowcasts based on the 10th percentile, $\tau=0.1$. For completeness, we also generate the nowcasts for the 50th ($\tau=0.5$) and 90th ($\tau=0.9$) percentiles to further investigate the behavior of the entire distribution of real GDP. The first type of nowcast is a standard forecast since the US real GDP has no release delay relative to the monthly variables under this category. In this case, both the monthly and quarterly US GDP variables have a balanced data structure, and no ragged edge occurs at the end of the sample. Thus, a growth-at-risk forecast is made for one to three months ahead.
The second and third types of nowcasts are produced from our model when the US real GDP is released with a one- and two-month delay, respectively. In both cases, the monthly and quarterly US GDP variables have an unbalanced data structure and a ragged edge at the end of the sample. Therefore, nowcasts of growth-at-risk are made under both categories.

\begin{figure}[t!h]
\centering
\hspace*{-3ex}
\includegraphics[width=16.0cm]{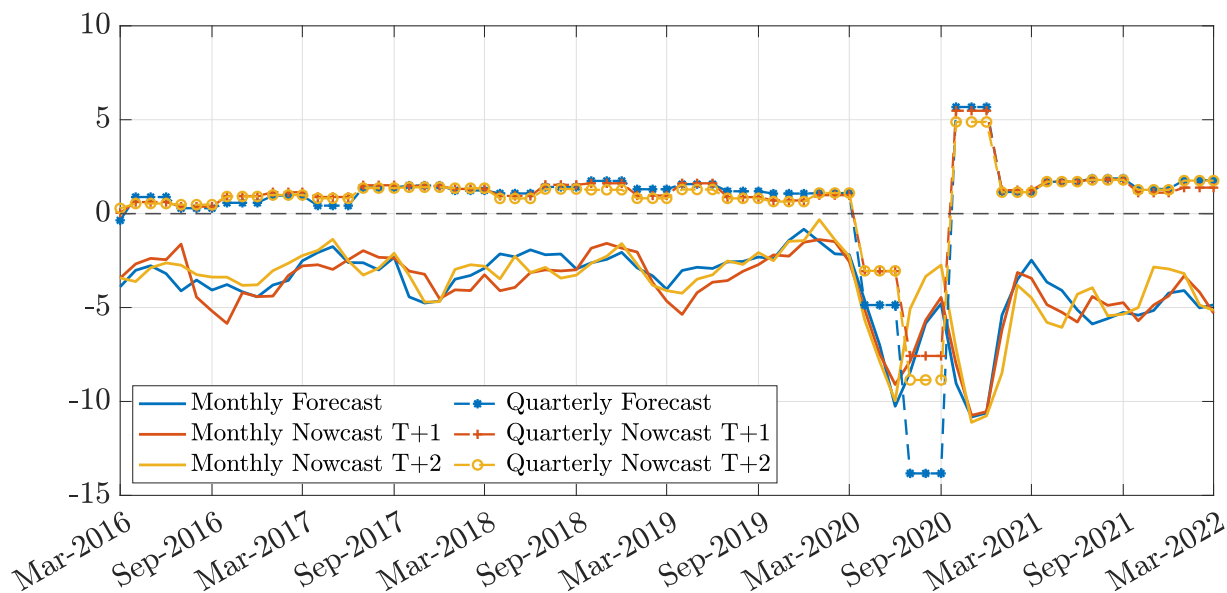}
\caption{Posterior mean of the rolling three-month average growth-at-risk changes
for the 10th percentile ($\tau=0.1$), from MF-QVAR (monthly, solid), and U-MIDAS QR (quarterly, dashed). Forecasts of $T$ in blue, nowcasts of $T+1$ in red, and nowcasts of $T+2$ in yellow.}
%Posterior mean of the rolling three-month average growth-at-risk changes for the 10th percentile ($\tau=0.1$).\\
%Notes: posterior estimates of the monthly growth-at-risk forecasts (blue), monthly growth-at-risk nowcasts of $T+1$ (red), and monthly growth-at-risk nowcasts of $T+2$ (yellow) from the MF-QVAR model.
%Quarterly forecasts (dashed, blue), quarterly nowcasts of $T+1$ (dashed, red), and quarterly nowcasts of $T+2$ (dashed, yellow) from the U-MIDAS QR model.}
\label{fig:postmomaverage}
\end{figure}

Figure~\ref{fig:postmomaverage} plots the rolling three-month posterior mean average of the monthly changes of the growth-at-risk estimates for the three types of nowcasts. It is evident from Figure~\ref{fig:postmomaverage} that the growth-at-risk nowcasts were, on average, about -3\% during the pre-pandemic period, whereas this average has fallen to about -5\% since the pandemic. This implies that COVID-19 has caused US monthly real GDP to be more skewed to the left and increased the vulnerability of the US to enter a recession.
In contrast, considering the 50th and 90th percentile, the monthly nowcast between spring and summer 2020 evolved in opposite direction compared to the quarterly predictions (see the Supplementary Material). This further highlights the importance of the proposed nowcasting approach in providing a timely characterization of risks.

Moreover, for comparison purposes, in Fig.~\ref{fig:postmomaverage} we also plot the corresponding quarterly growth-at-risk nowcasts from a U-MIDAS quantile regression (QR) model.
This U-MIDAS QR model extends the quarterly frequency QR model used in \cite{adrian2019vulnerable} and includes a monthly NFCI instead of a quarterly series utilized by \cite{adrian2019vulnerable}. The resulting quarterly growth-at-risk nowcasts are denoted as the dashed lines in Fig.~\ref{fig:postmomaverage}.
Most of these quarterly growth-at-risk nowcasts are positive, except for the initial year of the COVID-19 pandemic. In contrast, the monthly growth-at-risk nowcasts from the MF-QVAR model are all negative. These results suggest that the nowcasts from the U-MIDAS QR model may be underestimating the underlying growth-at-risk measure for US Real GDP.
For instance, the growth-at-risk nowcasts from the U-MIDAS QR model bounce back to their pre-pandemic level in 2021, which is inconsistent with recent global events. In fact, since the pandemic period, the US has experienced weaker growth and high inflation, and intuitively one would expect the US to be more prone to a recession than an expansion.
Conversely, the nowcasting results from our MF-QVAR model are consistent with this idea.

\begin{figure}[t!h]
\centering
\hspace*{-3ex}
\includegraphics[width=16.0cm]{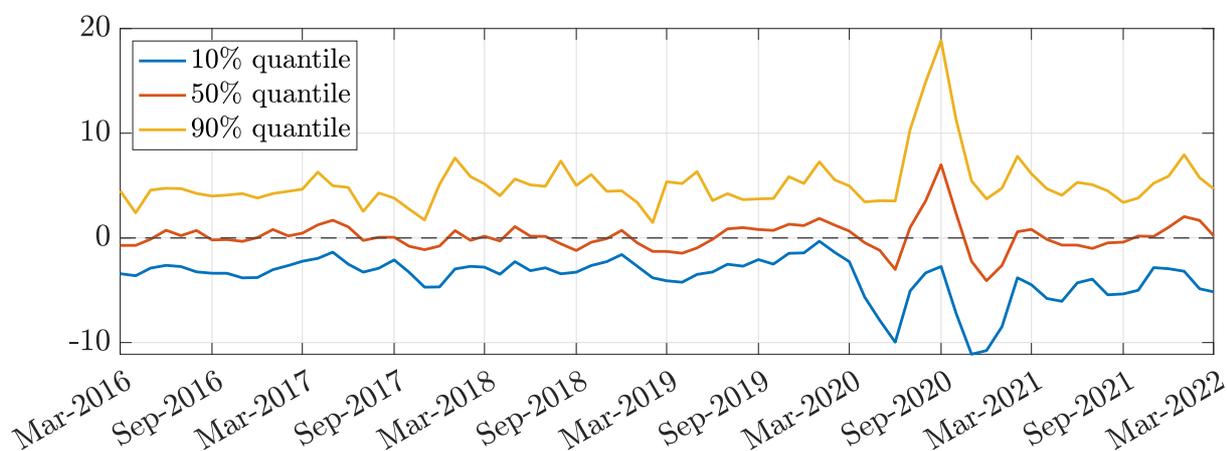}
\caption{Posterior mean of the rolling three-month average of the Monthly Nowcast $T+2$
for the 10th ($\tau = 0.1$, blue), 50th ($\tau = 0.5$, red), and 90th percentile ($\tau = 0.9$, yellow).}
%\\
%Notes: Posterior estimates of the Monthly Nowcast $T+2$ for the 10th (blue), 50th (red) and 90th (yellow) percentile, respectively.}
\label{fig:Monthly_Nowcast2}
\end{figure}

To further investigate the skewness of the real GDP nowcasts, we also generated the nowcasts for the 50th ($\tau=0.5$) and 90th ($\tau=0.9$) percentiles. Figure~\ref{fig:Monthly_Nowcast2} plots all the posterior estimates of the percentiles for the monthly nowcast of $T+2$. The pattern displayed in Fig.~\ref{fig:Monthly_Nowcast2} indeed confirms our previous finding that the COVID-19 pandemic has caused the real GDP nowcasts to become more negatively skewed.
Similar conclusions can be drawn from the monthly forecast and monthly nowcast of $T+1$ (see the Supplementary Material).
Moreover, in Table~\ref{tab:Nowcast2table} we report the posterior monthly nowcasts of all the percentiles for selected pandemic periods. For December 2019, the nowcast estimate for the 10th percentile was -1.44\%, dropping to -5.67\% in April 2020 of the first wave of the pandemic. In addition, the uncertainty associated with the real GDP nowcasts appears to have widened since the pandemic. For example, in Table~\ref{tab:Nowcast2table}, the distance between the nowcasts of the 10th and 90th percentile has significantly increased since December 2019. This result suggests that nowcasting real GDP has become inherently challenging since the pandemic.

\begin{table}[H]
\centering
\small
\begin{tabular}{ccccc}
\toprule 
 & \multicolumn{4}{c}{Percentile} \tabularnewline
\midrule 
Dates & $\tau=0.1$ & $\tau=0.5$ & $\tau=0.9$ & Difference between $\tau=0.9$ and $\tau=0.1$ \tabularnewline
\midrule
\midrule 
December 2019 & -1.44 & 1.17 & 5.20 & 6.63 \tabularnewline
\midrule 
April 2020 & -5.67 & -0.46 & 3.44 & 9.11 \tabularnewline
\midrule 
September 2020 & -2.74 & 6.99 & 18.84 & 21.58 \tabularnewline
\midrule 
January 2021 & -8.48 & -2.63 & 4.75 & 13.23 \tabularnewline
\midrule 
December 2021 & -2.95 & 1.01 & 5.90 & 8.85 \tabularnewline
\bottomrule
\end{tabular}
\caption{Posterior estimates of the 10th, 50th and 90th percentiles for the
monthly nowcast $T+1$ across selected periods during the pandemic
period.}
\label{tab:Nowcast2table}
\end{table}

\subsection{Does a tightening of the NFCI have a negative impact on US growth-at-risk?}
In this section, we conduct a counterfactual analysis to investigate the importance of NFCI in a nowcasting monthly US growth-at-risk. Specifically, we undertake the same real-time out-of-sample forecasting exercise described in the preceding section. In addition, for each window of nowcast made, we assume a tightening of the NFCI in the last three months, holding all other things constant. This will allow us to explicitly determine whether a tightening of the NFCI does indeed negatively impact monthly nowcasts of US growth-at-risk.

\begin{figure}[t!h]
\centering
\captionsetup{width=0.92\textwidth}
\hspace*{-4ex}
\begin{tabular}{cc}
\multicolumn{2}{c}{ \textbf{(a) Monthly Forecast} } \\
\multicolumn{2}{c}{
\includegraphics[trim= 0mm 0mm 0mm 0mm,clip,width= 0.52\textwidth]{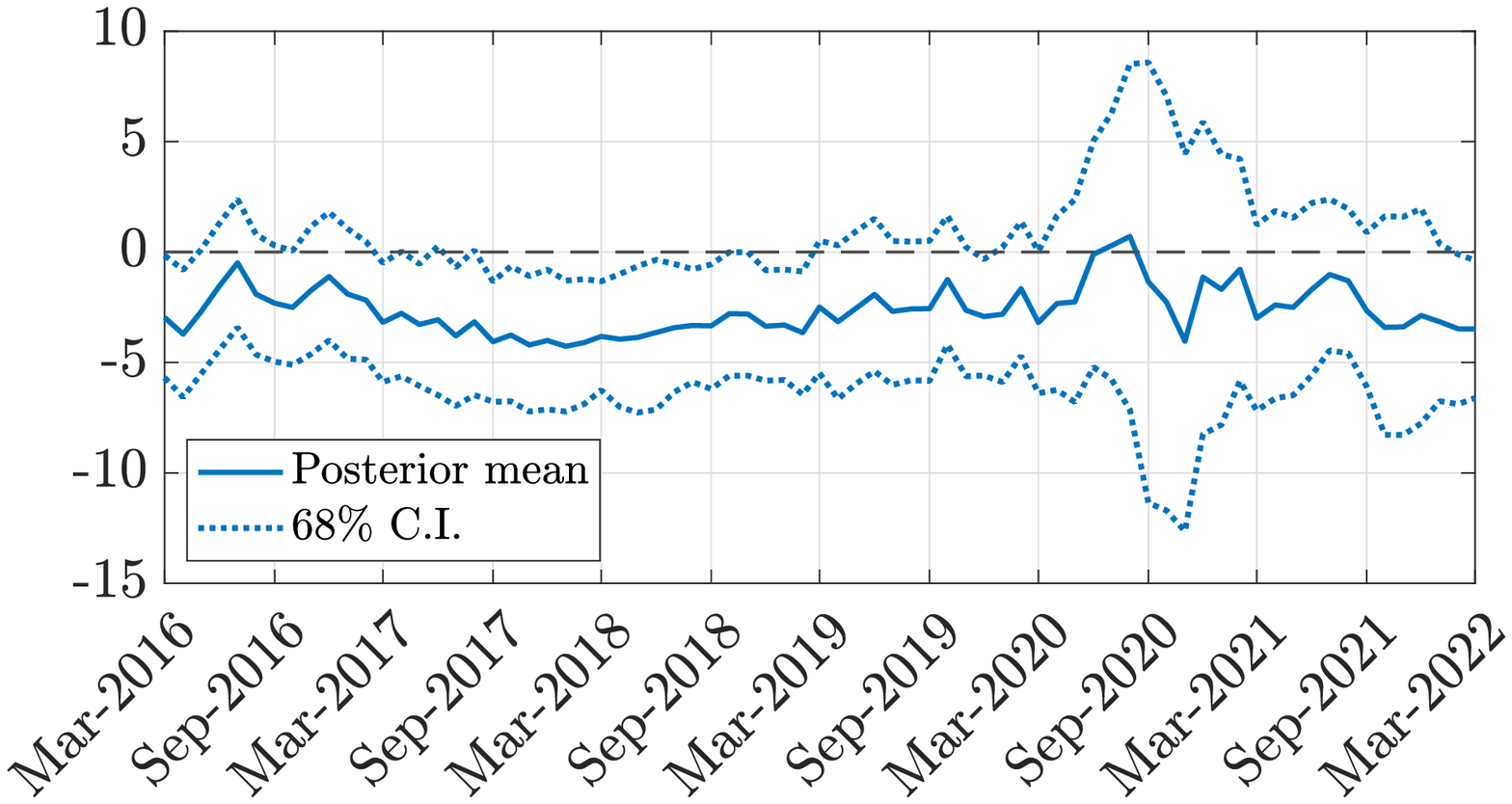}
} \\[7pt]
\textbf{(b) Monthly Nowcast $T+1$} & \textbf{(c) Monthly Nowcast $T+2$} \\
\includegraphics[trim= 0mm 0mm 0mm 0mm,clip,width= 0.52\textwidth]{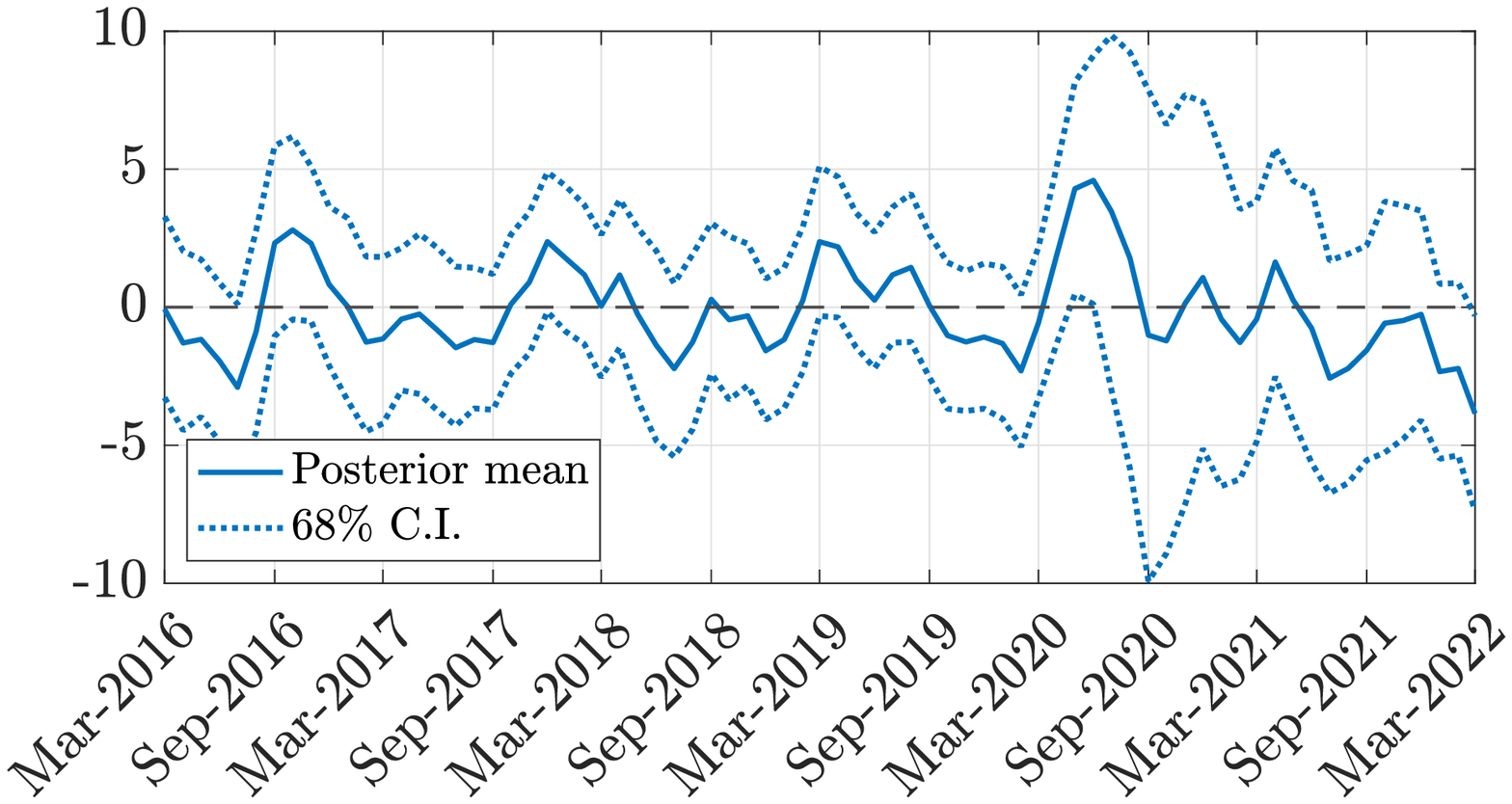} &
\includegraphics[trim= 0mm 0mm 0mm 0mm,clip,width= 0.52\textwidth]{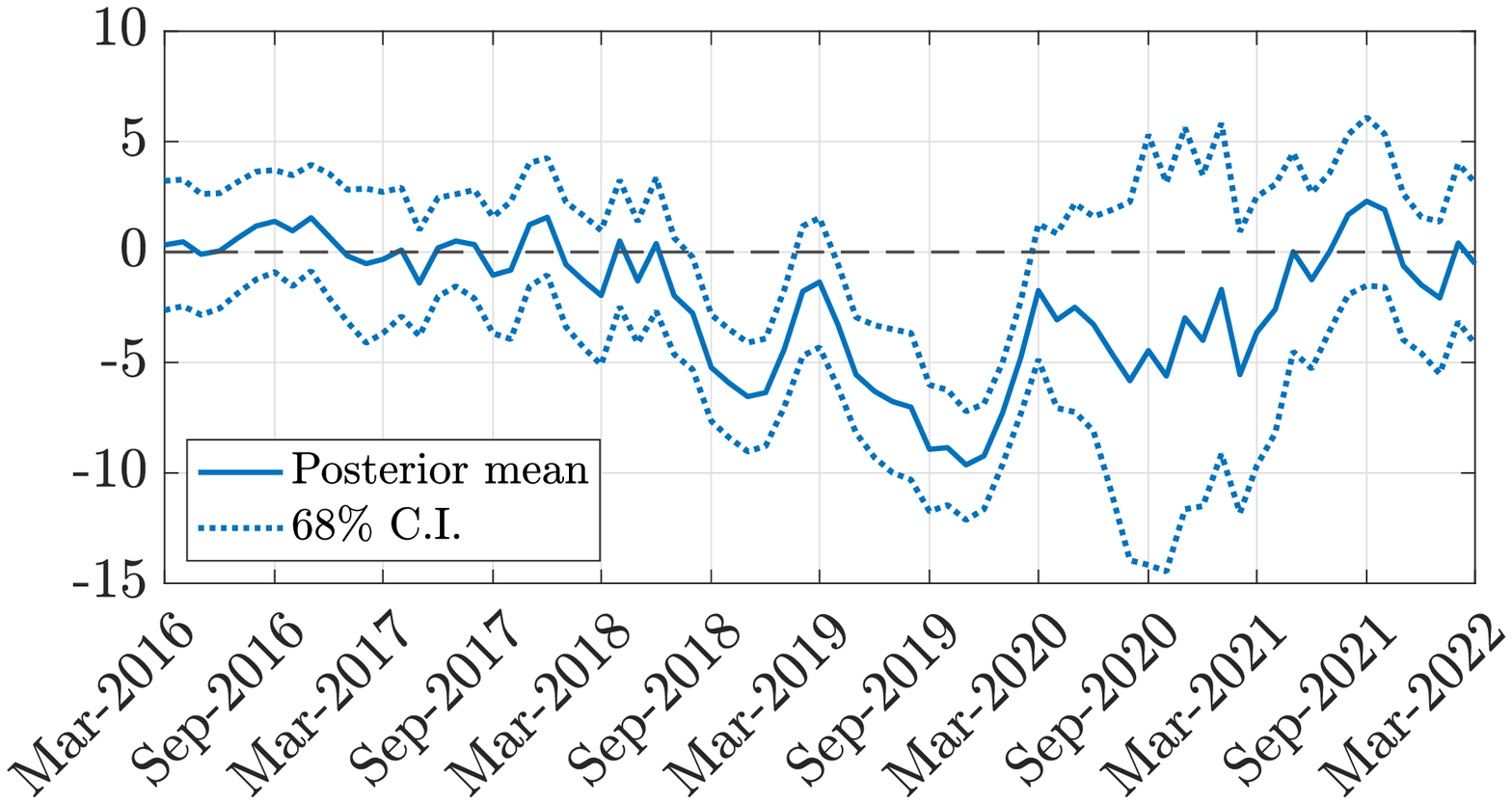}
%\textbf{(a) Monthly Forecast}\tabularnewline
%\includegraphics[trim= 15mm 5mm 15mm 5mm,clip,width= 0.7\textwidth]{\string"./figures/new/Monthly Forecast_counterfactual10percent\string".pdf}\tabularnewline
%\textbf{(b) Monthly Nowcast $T+1$}\tabularnewline
%\includegraphics[trim= 15mm 5mm 15mm 5mm,clip,width=0.7\textwidth]{\string"./figures/new/Monthly Nowcast T+1_counterfactual10percent\string".pdf}\tabularnewline
%\textbf{(c) Monthly Nowcast $T+2$}\tabularnewline
%\includegraphics[trim= 15mm 5mm 15mm 5mm,clip,width=0.7\textwidth]{\string"./figures/new/Monthly Nowcast T+2_counterfactual10percent\string".pdf}\tabularnewline
\end{tabular}
%\begin{tabular}{c}
%\textbf{(a) Monthly Forecast} \\
%\includegraphics[trim= 0mm 0mm 0mm 0mm,clip,width= 0.6\textwidth]{./figures/new/Monthly_Forecast_counterfactual10percent.eps} \\
%\textbf{(b) Monthly Nowcast $T+1$} \\
%\includegraphics[trim= 0mm 0mm 0mm 0mm,clip,width= 0.6\textwidth]{./figures/new/Monthly_Nowcast_T+1_counterfactual10percent.eps} \\
%\textbf{(c) Monthly Nowcast $T+2$} \\
%\includegraphics[trim= 0mm 0mm 0mm 0mm,clip,width= 0.6\textwidth]{./figures/new/Monthly_Nowcast_T+2_counterfactual10percent.eps}
%\end{tabular}
\caption{Posterior mean differences of the counterfactual and the actual real-time nowcasts (thick line) and the associated 68\% credible interval (dotted line), for time $T$, $T+1$, and $T+2$.}
\label{fig:counterfact}
\end{figure}

Figure~\ref{fig:counterfact} plots the posterior mean differences of the counterfactual and the actual real-time nowcasts for the three cases with their associated 68\% credible intervals. The posterior estimates were first calculated by taking the difference between each MCMC draw of the counterfactual and actual real-time nowcasts. Next, the average was computed across these differences. In general, a tightening of the NFCI appears to have a statistically significant negative impact on the monthly forecast and nowcast of  $T+2$. However, for the monthly nowcast $T+1$, a tightening of the NFCI appears to have a muted effect. Furthermore, Table~\ref{tab:averagediff} reports the averages of these differences over the evaluation period. A tightening of the NFCI appears to cause a deterioration in the monthly growth-at-risk forecast and nowcast of  $T+2$ by about 2-2.6\% on average.

These results infer that NFCI significantly impacts forecasts and nowcasts of monthly US growth-at-risk. In particular, NFCI significantly influences the nowcasts of growth-at-risk on months when US real GDP has a release delay of two months. Therefore, our results reinforce the \cite{adrian2019vulnerable} findings and further highlight the importance of including NFCI in a mixed frequency setting when modeling growth-at-risk.

\begin{table}[t!h]
\centering
\small
\begin{tabular}{c ccc}
\toprule 
Percentiles & \multicolumn{1}{c}{Forecast} & \multicolumn{1}{c}{Nowcast $T+1$} & \multicolumn{1}{c}{Nowcast $T+2$} \\
\midrule
\midrule 
10th Percentile & -2.63 & -0.13 & -2.23 \\
\midrule 
50th Percentile & -1.82 &  0.31 & -0.68 \\
\midrule 
90th Percentile &  1.35 &  1.32 &  2.04 \\
\bottomrule
\end{tabular}
\caption{The average posterior mean differences of the counterfactual and the actual real-time nowcasts across the out-of-sample evaluation period.}
\label{tab:averagediff}
\end{table}

\section{Conclusions}  \label{sec:conclusion}

Motivated by the limitations of popular VAR models for low-frequency economic variables, we introduce a novel mixed-frequency quantile vector autoregression (MF-QVAR) model.
The proposed method exploits the informational content of high- and low-frequency variables to produce forecasts and nowcasts of conditional quantiles for indicators of interest. This permits to derive quantile-related risk measures at high frequency, thus enabling timely policy interventions.

The MF-QVAR model admits a state-space representation where the measurement follows a multivariate asymmetric Laplace distribution. Bayesian inference is performed by means of an efficient MCMC algorithm that exploits a data augmentation scheme coupled with a precision sampler to estimate the missing low-frequency variables at higher frequencies.
%From a computational perspective, we build on the scale-mixture representation of the multivariate asymmetric Laplace distribution to design an efficient data-augmentation scheme in a Bayesian framework
%The informational content of low-frequency variables and the results from conditional mean models provide only limited evidence to investigate this problem.

The proposed method is applied to US macroeconomic data to obtain real-time nowcasts for the growth-at-risk of US real GDP. The results show the ability of MF-QVAR to produce meaningful monthly nowcasts that outperform the quarterly U-MIDAS QR benchmark and reveal interesting patterns at the outbreak and during the COVID-19 pandemic.
Moreover, a counterfactual analysis reveals that a contraction of NFCI has a negative and significant impact on the forecasts and nowcasts of monthly US growth-at-risk.

%Extensions of the method include the design of computationally efficient estimation in high dimensional QVAR models can be obtained, for example, by adopting a variational Bayes approach as in \cite{gefang2020computationally}.
%
%This article is concerned with small and medium size QVARs, however, computationally efficient estimation in high dimensional QVAR models can be obtained, for example, by adopting a variational Bayes approach as in \cite{gefang2020computationally}.

%%%%%%%%%%%%%%%%%%%%%%%%%%%%%%%%%%%%%%%%
%\bibliographystyle{agsm}
\bibliographystyle{chicago}
\bibliography{biblio}

\begin{thebibliography}{}

\bibitem[\protect\citeauthoryear{Adams, Adrian, Boyarchenko, and
  Giannone}{Adams et~al.}{2021}]{adams2021forecasting}
Adams, P.~A., T.~Adrian, N.~Boyarchenko, and D.~Giannone (2021).
\newblock Forecasting macroeconomic risks.
\newblock {\em International Journal of Forecasting\/}~{\em 37\/}(3),
  1173--1191.

\bibitem[\protect\citeauthoryear{Adrian, Boyarchenko, and Giannone}{Adrian
  et~al.}{2019}]{adrian2019vulnerable}
Adrian, T., N.~Boyarchenko, and D.~Giannone (2019).
\newblock Vulnerable growth.
\newblock {\em American Economic Review\/}~{\em 109\/}(4), 1263--89.

\bibitem[\protect\citeauthoryear{Berger, Morley, and Wong}{Berger
  et~al.}{2020}]{berger2020nowcasting}
Berger, T., J.~Morley, and B.~Wong (2020).
\newblock Nowcasting the output gap.
\newblock {\em Journal of Econometrics\/}.

\bibitem[\protect\citeauthoryear{Bernardi, Gayraud, and Petrella}{Bernardi
  et~al.}{2015}]{bernardi2015bayesian}
Bernardi, M., G.~Gayraud, and L.~Petrella (2015).
\newblock Bayesian tail risk interdependence using quantile regression.
\newblock {\em Bayesian Analysis\/}~{\em 10\/}(3), 553--603.

\bibitem[\protect\citeauthoryear{Bhadra, Datta, Polson, and Willard}{Bhadra
  et~al.}{2016}]{bhadra2016default}
Bhadra, A., J.~Datta, N.~G. Polson, and B.~Willard (2016).
\newblock Default {Bayesian} analysis with global-local shrinkage priors.
\newblock {\em Biometrika\/}~{\em 103\/}(4), 955--969.

\bibitem[\protect\citeauthoryear{Caggiano, Castelnuovo, and Nodari}{Caggiano
  et~al.}{2022}]{caggiano2017uncertainty}
Caggiano, G., E.~Castelnuovo, and G.~Nodari (2022).
\newblock Uncertainty and monetary policy in good and bad times: A replication
  of the vector autoregressive investigation by bloom (2009).
\newblock {\em Journal of Applied Econometrics\/}~{\em 37\/}(1), 210--217.

\bibitem[\protect\citeauthoryear{Casarin, Foroni, Marcellino, and
  Ravazzolo}{Casarin et~al.}{2018}]{casarin2018uncertainty}
Casarin, R., C.~Foroni, M.~Marcellino, and F.~Ravazzolo (2018).
\newblock Uncertainty through the lenses of a mixed-frequency {Bayesian} panel
  {Markov}-switching model.
\newblock {\em The Annals of Applied Statistics\/}~{\em 12\/}(4), 2559--2586.

\bibitem[\protect\citeauthoryear{Chan and Jeliazkov}{Chan and
  Jeliazkov}{2009}]{chan2009efficient}
Chan, J.~C. and I.~Jeliazkov (2009).
\newblock Efficient simulation and integrated likelihood estimation in state
  space models.
\newblock {\em International Journal of Mathematical Modelling and Numerical
  Optimisation\/}~{\em 1\/}(1-2), 101--120.

\bibitem[\protect\citeauthoryear{Chan, Koop, and Potter}{Chan
  et~al.}{2013}]{chan2013new}
Chan, J.~C., G.~Koop, and S.~M. Potter (2013).
\newblock A new model of trend inflation.
\newblock {\em Journal of Business \& Economic Statistics\/}~{\em 31\/}(1),
  94--106.

\bibitem[\protect\citeauthoryear{Chan, Poon, and Zhu}{Chan
  et~al.}{2021}]{chan2021efficient}
Chan, J.~C., A.~Poon, and D.~Zhu (2021).
\newblock Efficient estimation of state-space mixed-frequency {VARs}: {A}
  precision-based approach.
\newblock {\em arXiv preprint arXiv:2112.11315\/}.

\bibitem[\protect\citeauthoryear{Chavleishvili and Manganelli}{Chavleishvili
  and Manganelli}{2021}]{Manganelli2021quantileIRF}
Chavleishvili, S. and S.~Manganelli (2021).
\newblock Forecasting and {S}tress {T}esting with {Q}uantile {V}ector
  {A}utoregression.
\newblock Technical Report 2330, ECB Working Paper.

\bibitem[\protect\citeauthoryear{Cimadomo, Giannone, Lenza, Monti, and
  Sokol}{Cimadomo et~al.}{2021}]{cimadomo2021nowcasting}
Cimadomo, J., D.~Giannone, M.~Lenza, F.~Monti, and A.~Sokol (2021).
\newblock Nowcasting with large {Bayesian} vector autoregressions.
\newblock {\em Journal of Econometrics\/}.

\bibitem[\protect\citeauthoryear{Cong, Chen, and Zhou}{Cong
  et~al.}{2017}]{cong2017fast}
Cong, Y., B.~Chen, and M.~Zhou (2017).
\newblock Fast simulation of hyperplane-truncated multivariate normal
  distributions.
\newblock {\em Bayesian Analysis\/}~{\em 12\/}(4), 1017--1037.

\bibitem[\protect\citeauthoryear{Durbin and Koopman}{Durbin and
  Koopman}{2002}]{durbin2002simple}
Durbin, J. and S.~J. Koopman (2002).
\newblock A simple and efficient simulation smoother for state space time
  series analysis.
\newblock {\em Biometrika\/}~{\em 89\/}(3), 603--616.

\bibitem[\protect\citeauthoryear{Durbin and Koopman}{Durbin and
  Koopman}{2012}]{durbin2012time}
Durbin, J. and S.~J. Koopman (2012).
\newblock {\em Time series analysis by state space methods}, Volume~38.
\newblock OUP Oxford.

\bibitem[\protect\citeauthoryear{Ghysels}{Ghysels}{2016}]{ghysels2016macroeconomics}
Ghysels, E. (2016).
\newblock Macroeconomics and the reality of mixed frequency data.
\newblock {\em Journal of Econometrics\/}~{\em 193\/}(2), 294--314.

\bibitem[\protect\citeauthoryear{Ghysels, Santa-Clara, and Valkanov}{Ghysels
  et~al.}{2005}]{ghysels2005there}
Ghysels, E., P.~Santa-Clara, and R.~Valkanov (2005).
\newblock There is a risk-return trade-off after all.
\newblock {\em Journal of Financial Economics\/}~{\em 76\/}(3), 509--548.

\bibitem[\protect\citeauthoryear{Ghysels, Santa-Clara, and Valkanov}{Ghysels
  et~al.}{2006}]{ghysels2006predicting}
Ghysels, E., P.~Santa-Clara, and R.~Valkanov (2006).
\newblock Predicting volatility: getting the most out of return data sampled at
  different frequencies.
\newblock {\em Journal of Econometrics\/}~{\em 131\/}(1-2), 59--95.

\bibitem[\protect\citeauthoryear{Gneiting and Ranjan}{Gneiting and
  Ranjan}{2011}]{Gneiting2011qCRPS}
Gneiting, T. and R.~Ranjan (2011).
\newblock Comparing density forecasts using threshold- and quantile-weighted
  scoring rules.
\newblock {\em Journal of Business \& Economic Statistics\/}~{\em 29\/}(3),
  411--422.

\bibitem[\protect\citeauthoryear{Hauber and Schumacher}{Hauber and
  Schumacher}{2021}]{hauber2021precision}
Hauber, P. and C.~Schumacher (2021).
\newblock Precision-based sampling with missing observations: {A} factor model
  application.
\newblock {\em Deutsche Bundesbank Discussion Paper\/}.

\bibitem[\protect\citeauthoryear{Huber, Koop, Onorante, Pfarrhofer, and
  Schreiner}{Huber et~al.}{2020}]{huber2020nowcasting}
Huber, F., G.~Koop, L.~Onorante, M.~Pfarrhofer, and J.~Schreiner (2020).
\newblock Nowcasting in a pandemic using non-parametric mixed frequency {VARs}.
\newblock {\em Journal of Econometrics\/}.

\bibitem[\protect\citeauthoryear{Huber and Rossini}{Huber and
  Rossini}{2022}]{Huber2022BART}
Huber, F. and L.~Rossini (2022).
\newblock Inference in bayesian additive vector autoregressive tree models.
\newblock {\em The Annals of Applied Statistics\/}~{\em 16\/}(1), 104--123.

\bibitem[\protect\citeauthoryear{Hubrich and Tetlow}{Hubrich and
  Tetlow}{2015}]{hubrich2015financial}
Hubrich, K. and R.~J. Tetlow (2015).
\newblock Financial stress and economic dynamics: {The} transmission of crises.
\newblock {\em Journal of Monetary Economics\/}~{\em 70}, 100--115.

\bibitem[\protect\citeauthoryear{Kadiyala and Karlsson}{Kadiyala and
  Karlsson}{1997}]{kadiyala1997numerical}
Kadiyala, K.~R. and S.~Karlsson (1997).
\newblock Numerical methods for estimation and inference in bayesian
  var-models.
\newblock {\em Journal of Applied Econometrics\/}~{\em 12\/}(2), 99--132.

\bibitem[\protect\citeauthoryear{Kaufmann and Schumacher}{Kaufmann and
  Schumacher}{2019}]{kaufmann2019bayesian}
Kaufmann, S. and C.~Schumacher (2019).
\newblock Bayesian estimation of sparse dynamic factor models with
  order-independent and ex-post mode identification.
\newblock {\em Journal of Econometrics\/}~{\em 210\/}(1), 116--134.

\bibitem[\protect\citeauthoryear{Khalaf, Kichian, Saunders, and Voia}{Khalaf
  et~al.}{2021}]{khalaf2021dynamic}
Khalaf, L., M.~Kichian, C.~J. Saunders, and M.~Voia (2021).
\newblock Dynamic panels with {MIDAS} covariates: {Nonlinearity}, estimation
  and fit.
\newblock {\em Journal of Econometrics\/}~{\em 220\/}(2), 589--605.

\bibitem[\protect\citeauthoryear{Kilian and Vigfusson}{Kilian and
  Vigfusson}{2017}]{kilian2017role}
Kilian, L. and R.~J. Vigfusson (2017).
\newblock The role of oil price shocks in causing {US} recessions.
\newblock {\em Journal of Money, Credit and Banking\/}~{\em 49\/}(8),
  1747--1776.

\bibitem[\protect\citeauthoryear{Koenker and Bassett}{Koenker and
  Bassett}{1978}]{koenker1978regression}
Koenker, R. and G.~Bassett (1978).
\newblock Regression quantiles.
\newblock {\em Econometrica\/}, 33--50.

\bibitem[\protect\citeauthoryear{Koop, McIntyre, Mitchell, and Poon}{Koop
  et~al.}{2020}]{koop2020regional}
Koop, G., S.~McIntyre, J.~Mitchell, and A.~Poon (2020).
\newblock Regional output growth in the {United Kingdom}: {More} timely and
  higher frequency estimates from 1970.
\newblock {\em Journal of Applied Econometrics\/}~{\em 35\/}(2), 176--197.

\bibitem[\protect\citeauthoryear{Kotz, Kozubowski, and Podg{\'o}rski}{Kotz
  et~al.}{2001}]{kotz2001laplace}
Kotz, S., T.~Kozubowski, and K.~Podg{\'o}rski (2001).
\newblock {\em The Laplace distribution and generalizations: a revisit with
  applications to communications, economics, engineering, and finance}.
\newblock Number 183. Springer Science \& Business Media.

\bibitem[\protect\citeauthoryear{Mariano and Murasawa}{Mariano and
  Murasawa}{2003}]{mariano2003new}
Mariano, R.~S. and Y.~Murasawa (2003).
\newblock A new coincident index of business cycles based on monthly and
  quarterly series.
\newblock {\em Journal of Applied Econometrics\/}~{\em 18\/}(4), 427--443.

\bibitem[\protect\citeauthoryear{Mariano and Murasawa}{Mariano and
  Murasawa}{2010}]{mariano2010coincident}
Mariano, R.~S. and Y.~Murasawa (2010).
\newblock A coincident index, common factors, and monthly real {GDP}.
\newblock {\em Oxford Bulletin of Economics and Statistics\/}~{\em 72\/}(1),
  27--46.

\bibitem[\protect\citeauthoryear{Merlo, Petrella, and Raponi}{Merlo
  et~al.}{2021}]{Merlo2021VaR}
Merlo, L., L.~Petrella, and V.~Raponi (2021).
\newblock Forecasting var and es using a joint quantile regression and its
  implications in portfolio allocation.
\newblock {\em Journal of Banking \& Finance\/}~{\em 133}, 106248.

\bibitem[\protect\citeauthoryear{Mogliani and Simoni}{Mogliani and
  Simoni}{2021}]{mogliani2021bayesian}
Mogliani, M. and A.~Simoni (2021).
\newblock Bayesian {MIDAS} penalized regressions: {Estimation}, selection, and
  prediction.
\newblock {\em Journal of Econometrics\/}~{\em 222\/}(1), 833--860.

\bibitem[\protect\citeauthoryear{Montes-Rojas}{Montes-Rojas}{2019}]{MontesRojas2019MultQIRF}
Montes-Rojas, G. (2019).
\newblock Multivariate {Q}uantile {I}mpulse {R}esponse {F}uncitons.
\newblock {\em Journal of Time Series Analysis\/}~{\em 40}, 739--752.

\bibitem[\protect\citeauthoryear{Neal}{Neal}{2003}]{Neal2003slicesampling}
Neal, R.~M. (2003, 6).
\newblock Slice sampling.
\newblock {\em The Annals of Statistics\/}~{\em 31\/}(3), 705--767.

\bibitem[\protect\citeauthoryear{Petrella and Raponi}{Petrella and
  Raponi}{2019}]{petrella2019joint}
Petrella, L. and V.~Raponi (2019).
\newblock Joint estimation of conditional quantiles in multivariate linear
  regression models with an application to financial distress.
\newblock {\em Journal of Multivariate Analysis\/}~{\em 173}, 70--84.

\bibitem[\protect\citeauthoryear{Polson and Scott}{Polson and
  Scott}{2010}]{polson2010shrink}
Polson, N.~G. and J.~G. Scott (2010).
\newblock Shrink globally, act locally: {Sparse} {Bayesian} regularization and
  prediction.
\newblock {\em Bayesian Statistics\/}~{\em 9\/}(501-538), 105.

\bibitem[\protect\citeauthoryear{Rue}{Rue}{2001}]{rue2001fast}
Rue, H. (2001).
\newblock Fast sampling of {Gaussian} {Markov} random fields.
\newblock {\em Journal of the Royal Statistical Society: Series B (Statistical
  Methodology)\/}~{\em 63\/}(2), 325--338.

\bibitem[\protect\citeauthoryear{Rue and Held}{Rue and
  Held}{2005}]{rue2005gaussian}
Rue, H. and L.~Held (2005).
\newblock {\em Gaussian {Markov} random fields: {Theory} and applications}.
\newblock Chapman and Hall/CRC.

\bibitem[\protect\citeauthoryear{Schorfheide and Song}{Schorfheide and
  Song}{2015}]{schorfheide2015real}
Schorfheide, F. and D.~Song (2015).
\newblock Real-time forecasting with a mixed-frequency {VAR}.
\newblock {\em Journal of Business \& Economic Statistics\/}~{\em 33\/}(3),
  366--380.

\bibitem[\protect\citeauthoryear{Schorfheide, Song, and Yaron}{Schorfheide
  et~al.}{2018}]{schorfheide2018identifying}
Schorfheide, F., D.~Song, and A.~Yaron (2018).
\newblock Identifying long-run risks: {A} {Bayesian} mixed-frequency approach.
\newblock {\em Econometrica\/}~{\em 86\/}(2), 617--654.

\bibitem[\protect\citeauthoryear{Tian, Li, and Tian}{Tian
  et~al.}{2016}]{tian2016bayesian}
Tian, Y., E.~Li, and M.~Tian (2016).
\newblock Bayesian joint quantile regression for mixed effects models with
  censoring and errors in covariates.
\newblock {\em Computational Statistics\/}~{\em 31\/}(3), 1031--1057.

\bibitem[\protect\citeauthoryear{Tian, Tang, and Tian}{Tian
  et~al.}{2021}]{tian2021bayesian}
Tian, Y.-Z., M.-L. Tang, and M.-Z. Tian (2021).
\newblock Bayesian joint inference for multivariate quantile regression model
  with {$L_{1/2}$} penalty.
\newblock {\em Computational Statistics\/}~{\em 36\/}(4), 2967--2994.

\bibitem[\protect\citeauthoryear{Tobias and Brunnermeier}{Tobias and
  Brunnermeier}{2016}]{tobias2016covar}
Tobias, A. and M.~K. Brunnermeier (2016).
\newblock {CoVaR}.
\newblock {\em American Economic Review\/}~{\em 106\/}(7), 1705.

\bibitem[\protect\citeauthoryear{Yu and Moyeed}{Yu and
  Moyeed}{2001}]{yu2001bayesian}
Yu, K. and R.~A. Moyeed (2001).
\newblock Bayesian quantile regression.
\newblock {\em Statistics \& Probability Letters\/}~{\em 54\/}(4), 437--447.

\end{thebibliography}

\appendix

\section{Asymmetric Laplace distribution}   \label{sec:apdx_asymmetric_Laplace}

\paragraph{Univariate case.}
A random number $y\in\R$ follows an asymmetric Laplace distribution, denoted $y \sim \mathcal{AL}(\mu,\sigma,\tau)$, if
\begin{equation*}
f(y|\mu,\sigma,\tau) = \frac{\tau(1-\tau)}{\sigma} \exp\left\{ -\rho\left( \frac{y-\mu}{\sigma} \right) \right\},
\end{equation*}
where $\rho(\cdot)$ denotes the check loss function, $\sigma > 0$, $\mu \in\R$  and $\tau\in (0,1)$ are the scale, location, and skewness parameters, respectively.
The asymmetric Laplace distribution admits a representation as a location-scale mixture of Gaussian distributions, with exponentially-distributed mixing variable \citep{petrella2019joint,kotz2001laplace}, as follows:
\begin{align*}
y & = \mu + w \sigma \theta_{\tau,1} + \sigma \theta_{\tau,2} \sqrt{w} z, \qquad 
z \sim \mathcal{N}(0,1)  \quad  w \sim \mathcal{E}xp(1),
\end{align*}
where the rate parametrization of the exponential distribution is used\footnote{Using the rate parametrization of the exponential distribution, it holds $k \mathcal{E}xp(\sigma) \stackrel{d}{=} \mathcal{E}xp(\sigma/k)$.} and
\begin{equation*}
\theta_{\tau,1} = \frac{1-2\tau}{\tau(1-\tau)}, \qquad 
\theta_{\tau,2} = \sqrt{\frac{2}{\tau(1-\tau)}}.
\end{equation*}
Therefore, it holds:
\begin{align*}
f(y|\mu,\sigma,\tau) = \bigintsss_{\, 0}^{+\infty} \exp\{-w\} \big( 2\pi \sigma^{2} \theta_{\tau,2}^{2} w \big)^{-\frac{1}{2}} \exp\left\{ -\frac{(y - \mu- \sigma \theta_{\tau,1} w)}{2 \sigma^{2}\theta_{\tau,2}^{2} w} \right\} \: dw.
\end{align*}

\paragraph{Multivariate case.}
An $n$-dimensional random vector $\mathbf{y}\in\R^n$ follows a multivariate asymmetric Laplace distribution, denoted $\mathbf{y} \sim \MAL_n(\boldsymbol{\mu}, D\boldsymbol{\theta}_{\tau,1}, D\btheta_{\tau,2} \Psi \btheta_{\tau,2} D)$, if
\begin{align*}
\mathbf{y} & = \boldsymbol{\mu} + w D \boldsymbol{\theta}_{\tau,1} + \sqrt{w} D \btheta_{\tau,2}^{1/2} \Psi^{1/2} \mathbf{z}, \qquad w \sim \mathcal{E}xp(1), \quad \mathbf{z} \sim \mathcal{N}_n(0,I_n),
\end{align*}
where $\Psi$ is a correlation matrix, $D = \operatorname{diag}(\delta_{1}, \ldots, \delta_{n})$, and 
\begin{align*}
\btheta_{\tau,1}  = \left( \frac{1-2\tau_{1}}{\tau_{1}(1-\tau_{1})}, \ldots, \frac{1-2\tau_{n}}{\tau_{n}(1-\tau_{n})} \right)', \qquad
\btheta_{\tau,2}  = \operatorname{diag} \left( \sqrt{\frac{2}{\tau_{1}(1-\tau_{1})}}, \ldots, \sqrt{\frac{2}{\tau_{n}(1-\tau_{n})}} \ \right).
\end{align*}
This implies that each entry $y_i$ is given by
\begin{align*}
y_i & = \mu_i + w \delta_i \theta_{\tau,1,i} + \sqrt{w} \delta_i \theta_{\tau,2,i} z_i, \quad w \sim \mathcal{E}xp(1), \quad z_i \sim \mathcal{N}(0,1),
\end{align*}
thus providing
\begin{align*}
y_i \sim \mathcal{AL}(\mu_i,\delta_i,\tau_i).
\end{align*}

\end{document}